\pdfoutput=1
\documentclass[amsmath,amssymb,aps,twocolumn,superscriptaddress]{revtex4-2}
\usepackage{amsmath}
\usepackage{amssymb}
\usepackage{bm}
\usepackage{graphicx}
\usepackage{epsf}
\usepackage{xcolor}
\usepackage{color}
\usepackage{graphicx}% Include figure files
\usepackage[version=4]{mhchem}% change $\mathrm{SO_2}$ or SiO${_2}$ to -> \ce{SiO2}
\usepackage[bookmarks=false]{hyperref}
\usepackage{siunitx}% try to write down numbers and units in a consistent way
\usepackage[english]{babel}
\usepackage[T1]{fontenc}
\usepackage{dcolumn}% Align table columns on decimal point
\usepackage{tabularx}
\usepackage{array} 
\usepackage[english]{babel}
\usepackage{xcolor}

\makeatletter
\def\maketitle{
\@author@finish
\title@column\titleblock@produce
\suppressfloats[t]}
\makeatother
\newcommand{\beginsupplement}{
    \setcounter{table}{0}
    \renewcommand{\thetable}{S\arabic{table}}
    \setcounter{figure}{0}
    \renewcommand{\thefigure}{S\arabic{figure}}
    \setcounter{equation}{0}
    \setcounter{section}{0}
    \renewcommand{\theequation}{S\arabic{equation}}
}

\begin{document}
\preprint{APS/123-QED}

\title{\textbf{Mirror-mediated long-range coupling and robust phase locking of spatially separated exciton-polariton condensates}
}
\author{Shuang Liang}
\thanks{Contact author: shl450@pitt.edu}
\affiliation{Department of Physics, University of Pittsburgh, 3941 O’Hara Street, Pittsburgh, Pennsylvania 15218, USA}

\author{Hassan Alnatah}
%\affiliation{Department of Physics, University of Pittsburgh, 3941 O’Hara Street, Pittsburgh, Pennsylvania 15218, USA}
\altaffiliation{H.A. and S.L. contributed equally to this work}
\affiliation{Department of Physics, University of Pittsburgh, 3941 O’Hara Street, Pittsburgh, Pennsylvania 15218, USA}

\author{Qi Yao}
\affiliation{Department of Physics, University of Maryland, College Park, Maryland, 20742, USA}

\author{Jonathan Beaumariage}
\affiliation{Department of Physics, University of Pittsburgh, 3941 O’Hara Street, Pittsburgh, Pennsylvania 15218, USA}

\author{Ken West}
\affiliation{Department of Electrical Engineering, Princeton University, Princeton, New Jersey 08544, USA}

\author{Kirk Baldwin}
\affiliation{Department of Electrical Engineering, Princeton University, Princeton, New Jersey 08544, USA}

\author{Adbhut Gupta}
\affiliation{Department of Electrical Engineering, Princeton University, Princeton, New Jersey 08544, USA}

\author{Loren N. Pfeiffer}
\affiliation{Department of Electrical Engineering, Princeton University, Princeton, New Jersey 08544, USA}

\author{Natalia G. Berloff}
\affiliation{Department of Applied Mathematics and Theoretical Physics, University of Cambridge, Cambridge CB3 0WA, United Kingdom}

\author{David W. Snoke}
\affiliation{Department of Physics, University of Pittsburgh, 3941 O’Hara Street, Pittsburgh, Pennsylvania 15218, USA}

\date{\today}

\begin{abstract}

Lattice arrays are valuable simulators for complex mathematical problems, but physical systems typically allow only short-range coupling. We demonstrate a method for independently tunable, long-range interactions between polariton condensates in two-dimensional lattices by using vertical emission and external imaging to couple arbitrary sites. Two geometrically isolated condensates are phase-locked without planar coupling, verified via phase-resolved interferometry. Analytical modeling reveals mechanisms for robust coherence. The mirror-based scheme, free of cameras or modulators, offers a pure, high-bandwidth analogt element. Extension to dense graphs via segmented micro-mirrors is limited only by imaging optics, enabling scalable, energy-efficient polaritonic hardware for neuromorphic computation.

\end{abstract}

\maketitle

%%%%%%%%%%%%%%%%%%%%%%%%%%%%%%%%%%%%%%%%%%%%%%%%%%%%%%
%%%%%%%%%%%%%%%%%%%%%%%%%%%%%%%%%%%%%%%%%%%%%%%%%%%%%%
\section{INTRODUCTION}

Physical neural networks (PNNs) solve optimization and learning tasks by allowing a dynamical system to relax to a low-energy configuration that encodes the answer.  Recent hardware realizations such as coherent Ising machines built from time-multiplexed optical parametric oscillators\,\cite{marandi2014network}, networks of injection-locked lasers\,\cite{Nixon2013ogf}, CMOS digital annealers\,\cite{sao2019application}, coupled electronic oscillators\,\cite{chou2019analog}, simulated-bifurcation processors\,\cite{tatsumura2019fpga}, analog iterative machines \cite{kalinin2023analog}, probabilistic Ising machines \cite{aadit2022massively} and SLM-based photonic Ising machines\,\cite{pierangeli2019large, veraldi2025fully} already rival digital heuristics on benchmark Ising and XY problems.  Their shared advantage is massively parallel, low-power evaluation of cost functions that encode a wide range of NP-hard tasks via polynomial reduction to the Ising or XY models\,\cite{lucas2014ising, wang2025phase}.  These developments, in turn, motivate new classical algorithms\,\cite{kalinin2018networks,leleu2019destabilization, kamaletdinov2024coupling} and benchmark problems\,\cite{hamze2020wishart,stroev2024benchmarking}.

Quantum, photonic and hybrid light–matter PNNs promise still greater gains\,\cite{Stroev2023Analog,syed2023beyond}.  Photonic Ising engines achieve $\mathcal O(N)$ runtime for low-rank interaction matrices\,\cite{pierangeli2019large,pierangeli2020adiabatic}, and focal-plane partitioning has recently broadened their scope\,\cite{veraldi2025fully}.  Gain-based condensate platforms such as  exciton–polaritons or QEDs convert optimisation tasks into the gain/loss landscape of a driven system, selecting the minimum-loss mode at threshold\,\cite{Berloff2017,marsh2023entanglement, Stroev2023Analog}.  A decisive obstacle to scaling PNNs beyond proof-of-concept lattices is the lack of fast, fully programmable long-range couplings. Quantum annealers struggle to distribute entanglement over sparse qubit graphs, spatial-photonic Ising machines (SPIMs) resort to low-rank factorisations to emulate the full $N^{2}$ interaction matrix, and polariton lattices restricted to reservoir overlap are confined to near-neighbour links that cannot be tuned pairwise. Without dense, adjustable connectivity one cannot tackle large combinatorial instances or train analogue neural networks, whose synaptic weights must be updated iteratively and often non-locally. An adequate solution must extend well beyond geometric overlap, allow element-by-element programming of arbitrary coupling 
  matrices, and avoid the slow electronic feedback loops that would erase the picosecond advantage of exciton-polariton dynamics. 
  
  We meet these criteria by introducing mirror-mediated photon feedback: a passive high-reflectivity mirror retro-injects leakage light, establishing centimetre-range, sign-reversible, continuously tuneable links between otherwise isolated condensates. This mechanism supplies precisely the  reconfigurable coupling structure required for scalable optimisation and neuromorphic learning in polaritonic hardware. Crucially, this is the first step in creating  the hardware that acts as a pure polaritonic optimizer: all couplings pre‑programmed optically with no run‑time electronic intervention, where a fast SLM can update the couplings between iterations.  The pure regime preserves the fundamental speed and energy advantages of polaritons.

Exciton–polaritons are formed by strong coupling of cavity photons to quantum-well excitons\,\cite{kasprzak2006bose}.  With a pump just above threshold, a macroscopic population collapses into the lowest-loss collective mode on a sub-picosecond time-scale.  In a lattice of $N$ traps the order parameter $\psi_j=\rho_j e^{i\theta_j}$ obeys a driven–dissipative Gross–Pitaevskii equation (complex Ginzburg-Landau equation, cGLE) whose phase degree of freedom  reduces to the XY Hamiltonian when the amplitudes $\rho_j$ are about the same \cite{Berloff2017}
\begin{equation}
H_{\text{XY}}=-\sum_{i<j}J_{ij}\cos(\theta_i-\theta_j).
\end{equation}
The amplitudes can be made exactly the same at the equilibrium point by providing the feedback on the injection strength \cite{kalinin2018networks}.
A binary-phase resonant drive $\theta_i\in\{0,\pi\}$ converts the same hardware into an Ising machine\,\cite{kalinin2018simulating}.  Fabrication  based on optical imprinting\,\cite{wertz2010spontaneous,manni2011spontaneous,tosi2012sculpting,tosi2012geometrically}, patterned perovskites\,\cite{su2018room, peng2024room}, etched mesas\,\cite{winkler2015polariton}, strain traps\,\cite{balili2007bose}, surface-acoustic waves\,\cite{cerda2012dynamic} and open micro-cavities\,\cite{dufferwiel2014strong} yield lattices of tens to hundreds of sites, yet their native geometric coupling is short-ranged and includes coherent Josephson terms that change the energy landscape\,\cite{kalinin2019polaritonic}.  The ways to introduce the controllable pairwise interactions were theoretically introduced \cite{kalinin2020polaritonic} but were never before experimentally realized.

We, therefore, isolate condensates geometrically and introduce interactions by retro-injecting leakage photons with a planar mirror.  The mirror reflects emission from condensate $j$ back into condensate $i$, adding a  cross-term whose magnitude grows with mirror reflectivity, while the change in the round-trip optical path switches its sign.  The round-trip optical path is of order of 10 cm and introduces feedback delay of $\tau$.  Interferometry confirms deterministic phase locking between the condensates as we show below.

Replacing the single mirror with a segmented micro-mirror array would enable an arbitrary $J_{ij}$ matrix by directing separately addressable feedback spots to multiple condensates.  The practical limit is set by the field of view and numerical aperture of the imaging optics: as the number of spots grows, diffraction, aberrations and mechanical cross-talk between mirror facets will eventually degrade the feedback phase fidelity and reduce $|J_{ij}|$.

\section{EXPERIMENTAL METHODS}
In the experiments reported here, we used a GaAs/AlGaAs microcavity structure very similar to those of previous experiments \cite{alnatah2025strong,alnatah2024critical,alnatah2024bose}.  The microcavity sample consisted of a total of 12 GaAs quantum wells with AlAs barriers embedded within a distributed Bragg reflector (DBR). The quantum wells are in groups of four, with each group placed at one of the antinodes of cavity. A key feature of the experiments described here is that a sample with lower $Q$-factor was used than the best possible. For the very high-$Q$ samples used in previous works (e.g., \cite{steger2015slow,nelsen2013dissipationless,alnatah2024coherence}), the reflectivity of the top DBR is so high that very little external light can get in. We therefore used a variation with $Q$ nominally equal to $\sim 4\times 10^4$.
\par
The polaritons were generated by pumping the sample non-resonantly with a wavelength-tunable laser, tuned to a reflectivity minimum approximately 103 meV above the lower polariton resonance (771.5 nm). The laser profile was shaped using a a spatial light modulator (SLM) to create six circular spots as shown in Fig.~\ref{fig1}(a). The pump laser served as an incoherent source of excitons and polaritons because its photon energy was well above the exciton resonance. As a result, hot carriers were generated that required multiple scatterings to cool before forming excitons and, subsequently, polaritons. This incoherent excitation had two key effects: (1) it created a repulsive potential landscape due to the accumulation of slow-moving excitons in the pump region, which acted as a trap for polaritons, (2) and it also acted as the source of the polaritons, as excitons converted into polaritons. To minimize heating of the sample, the pump laser was modulated using an optical chopper with a duty cycle of 1.7$\%$ and pulses of duration approximately 41.6 $\mathrm{\mu s}$, which is very long compared to the dynamics of the system. The non-resonant pump created electrons and holes, which scattered down in energy to become polaritons.  The photoluminescence (PL) was collected using a microscope objective with a numerical aperture of 0.75 and was imaged onto the entrance slit of a spectrometer. The image was then sent through the spectrometer to a CCD camera for time-integrated imaging. In all the experiments described here, the sample was held in a cryostat at low temperature (5 K).
\par
As in many previous studies \cite{deng2002condensation,kasprzak2006bose,balili2007bose,abbarchi2013macroscopic,sanvitto2010persistent,lagoudakis2009observation}, we can easily switch between real-space imaging and momentum-space imaging by a simple change of the focal plane. Both of these can be done with energy resolution by sending the light through an imaging spectrometer.

\section{EXPERIMENTAL RESULTS}

Proving that two condensates are coupled only through an external path is tricky, because when an imaging interferometer is used, there are several spots: two condensates, two spots corresponding to the light from each condensate sent to the other through the external path, and two copies of all four of these spots sent through the two legs of the interferometer. We must therefore be careful to establish the logic of the experiment.

\begin{figure}[htbp]
  \centering
  \includegraphics[width=0.5\textwidth]{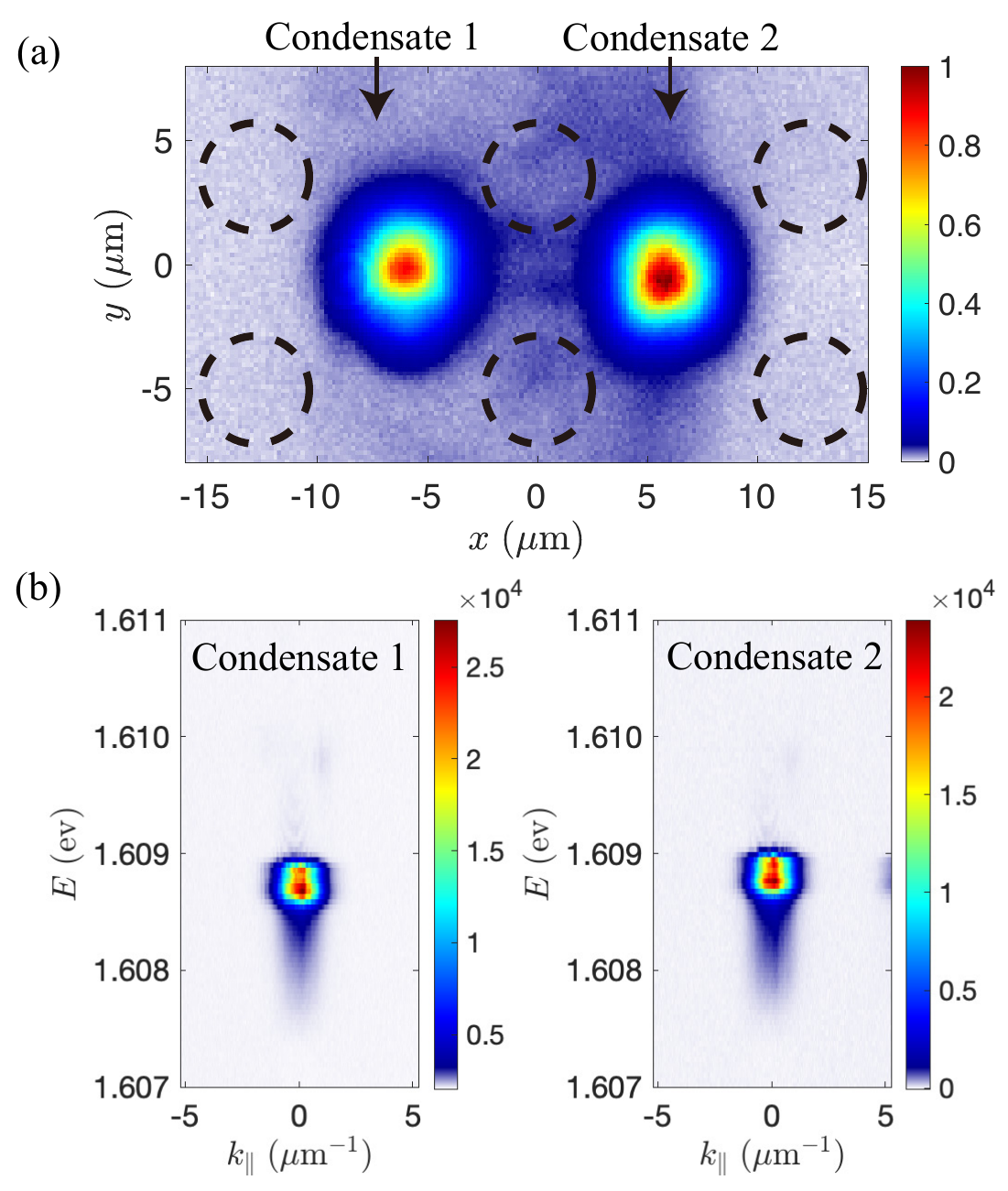}
      \centering
  \caption{{\bf Two independent condensates}. (a) Real-space image. The dashed circles indicate the pump laser positions. The excitation power is $P = 1.5\,P_\mathrm{th}$. The threshold power $P_\mathrm{th}$ is defined in the Supplementary Material. (b) Energy vs. in-plane wave vector image of the two condensates under the same conditions. Each has energy profile given by the spectral resolution of the imaging spectrometer used. }
  \label{fig1}
\end{figure}

We prove that the two condensates are coupled coherently only via the external feedback by the following set of steps:
\begin{enumerate}
\item {\bf Two independent condensates are created, each in a potential-energy minimum, confined by four neighboring exciton clouds.} Figure~\ref{fig1}(a) shows the real-space image of the emission from the two condensates. The dashed circles show the locations of the pump laser spots that create exciton barriers. The energy of the two condensates is tuned to be exactly equal, within our spectral resolution, by changing the spacing of the confining exciton clouds, which can be used to shift the confinement energy of the two traps independently. Figure~\ref{fig1}(b) shows the measured energy spectrum of the PL from each condensate. 

\item {\bf The two condensates are shown to not have in-plane coupling, by overlapping the light from each through a Michelson interferometer.} Figure~\ref{fig2}(a) shows the layout of the interferometer, and Figure~\ref{fig2}(b) shows the interference pattern when the two images of the condensates are flipped left-to-right and overlapped. There is a weak interference in the middle region between the two, from leaking condensate that has escaped the traps, but no interference of the light from the trapped condensates.

\begin{figure}[htbp]
  \centering
  \includegraphics[width=0.5\textwidth]{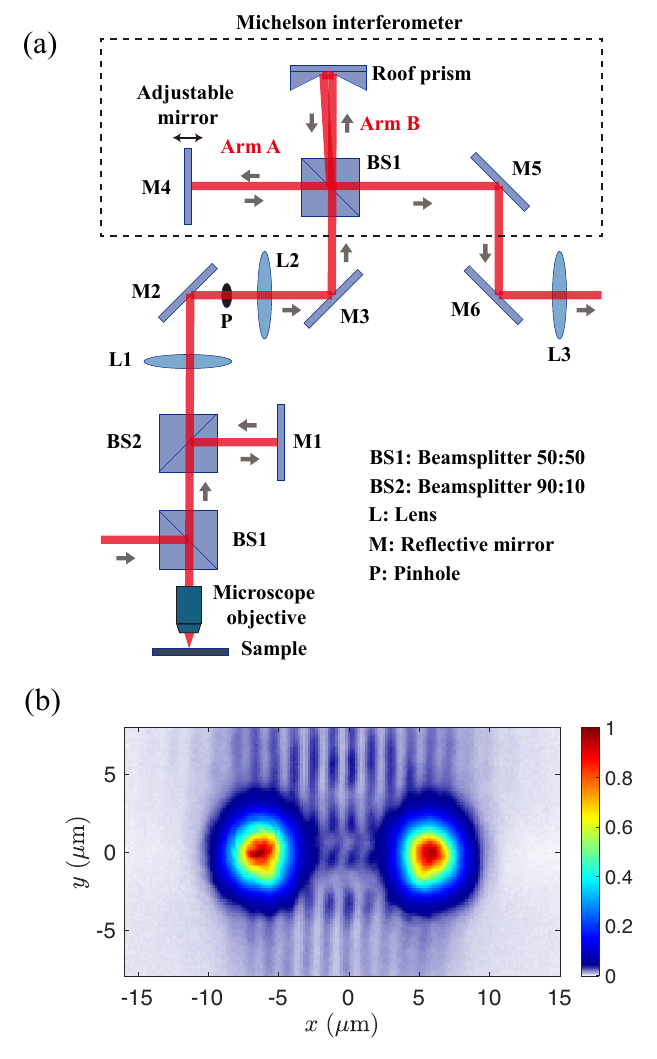}

  \centering
  \caption{(a) The layout of the optical setup. (b) The interference pattern
when the two images of the condensates are flipped
left-to-right and overlapped. }
  \label{fig2}
\end{figure}

\item {\bf Light emission from each condensate is collected by a secondary, external imaging system and refocused onto a spot near the other condensate.} The feedback path of this light hits mirror M1 in the layout shown in Figure~\ref{fig2}(a), and returns through the same microscope objective.
Figure~\ref{fig3}(a) shows the image of Condensate 1 and its feedback spot, when only this condensate has been created, and Figure~\ref{fig3}(b) shows the image of Condensate 2 and its feedback spot, when only this condensate has been created. No interferometer was used for these images. 

\begin{figure}[htbp]
  \centering
  \includegraphics[width=0.45\textwidth]{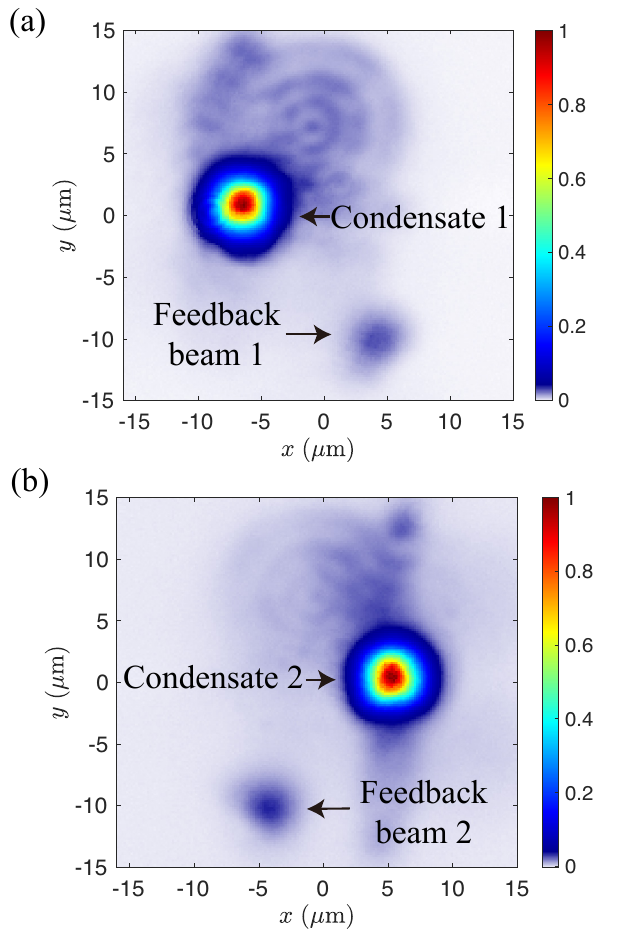}
  \caption{(a) Real-space image of Condensate 1 and its feedback beam. No interferometry is used. (b) The same for Condensate 2. }
  \label{fig3}
\end{figure}

\item {\bf The path lengths of the two arms of the Michelson interferometer are adjusted so that the light from one condensate is interfered with the light from the other condensate after the proper delay time.}   The feedback light follows a lengthy additional path, about 47 cm, out through the imaging system and then back to the sample; the travel time for this path is longer than the condensate coherence time (~141 ps), as determined from the measurements presented in Fig. S5 of the supplementary material. Therefore one leg of the Michelson interferometer is adjusted to account for this time delay, as illustrated in Figure~\ref{fig2}(a).  Importantly, this “delay compensation” means the interferometer is not probing equal‑time coherence between fields at the sample. Rather, it probes a time‑shifted first‑order correlation: the field emitted at time $t-\tau$ is interfered with the field at time t, where $\tau$ is the external feedback propagation delay. Consequently, observing fringes at the compensated setting does not imply that coherence persists for times $\tau$ ; it implies that the relevant correlation function has a peak (revival) at the specific delay $\tau$ . If the interferometer delay is detuned from this compensation by $\Delta=\tau + \delta$, the fringe visibility is expected to decrease as $|\delta|$ exceeds the intrinsic single‑condensate coherence time. Figure~\ref{fig4} shows the interference of one condensate with its own feedback light after it has completed the round trip back to the sample, without the other condensate present, showing that the delay time has been adjusted properly. 

Note that there is now an asymmetry between the two interference images. The image on left side corresponds to the overlap of the image of Condensate 1 with the feedback light when both signals have traveled the same distance. The image on the right side shows the same two spots, but these do not travel equal distances on both paths, and therefore there is no interference, since the different in paths is greater than the coherence time of the condensate.

\begin{figure}[htbp]
  \centering
  \includegraphics[width=0.44\textwidth]{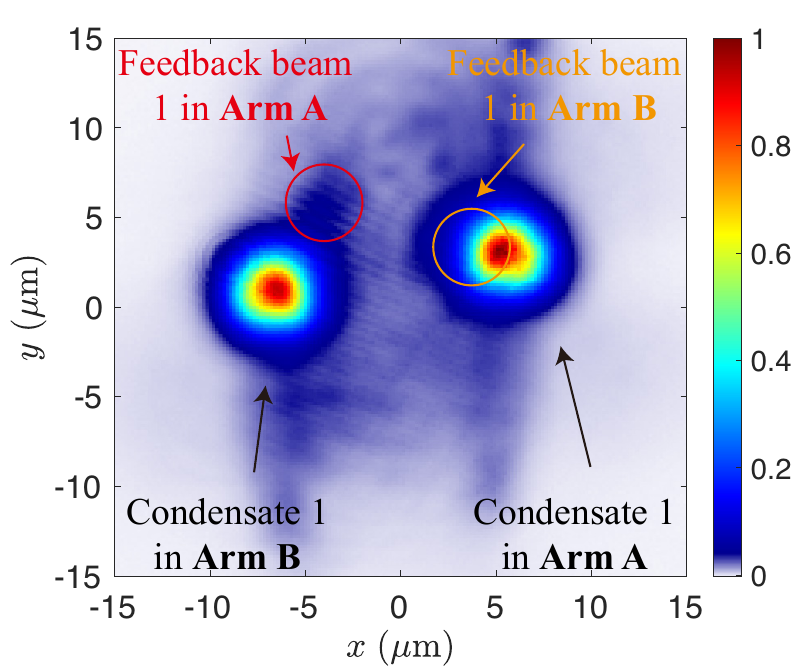}
  \caption{Interference of the image of Condensate 1 and its own feedback beam through Arm A, with the same image reversed, through Arm B. The feedback beam has been moved to a new position compared to its location in Figure~\ref{fig3}. Note that there are fringes only for the condensate light in Arm B with the feedback light in Arm A, because only this case has the light from the two paths with equal time delay.  }
  \label{fig4}
\end{figure}

\item {\bf With the feedback light from one condensate going onto the other, interference shows that the two condensates now are phase locked.} Figure \ref{fig5}(a) shows the image from the same experimental configuration as Figure \ref{fig4}, but now with both condensates (and their feedback beams) present. Fringes are seen across the entire condensate area. As seen in Figure~\ref{fig5}(c), the fringe visibility is about 10\%. 
Here ``phase locked'' is meant in the delay-synchronization (time-lag locking) sense.
Because condensate~2 is driven by the \emph{delayed} injected field originating from condensate~1, the locked relation is between phases at different times, schematically
\begin{equation}
\theta_2(t)\approx \theta_1(t-\tau)+\theta_0,
\label{eq:timelag_locking}
\end{equation}
(and analogously for mutual coupling).
The reduced visibility compared to the self-interference case indicates partial locking and residual relative-phase fluctuations; it does not imply nanosecond-scale equal-time mutual coherence.
 Fluctuations of the condensates may be responsible for less than perfect coherence, but there is clearly a phase correlation of the two.

In principle, the feedback light from Condensate 1 hitting Condensate 2 can also interfere with the image of Condensate 1. To avoid this, the feedback spot was moved to one side of condensate 2, and the overlap of the images of condensates 1 and 2 included only a tiny fraction of this spot. Therefore, the only way for there to be interference of the two images is if the feedback seed from Condensate 1 brings Condensate 2 into phase locking as a whole entity, and then the far side of that Condensate 2 interferes with the light from Condensate 1. To verify that the interference between a condensate and its own feedback beam does not influence the phase-locking observations, we extracted the interference fringes from Fig. 4. The results are shown in Fig. S3 and discussed in the supplementary material. We have also indicated the position of feedback beam 2 in Fig. S6 of the supplementary material.

\begin{figure}[htbp]
  \centering
  \includegraphics[width=0.44\textwidth]{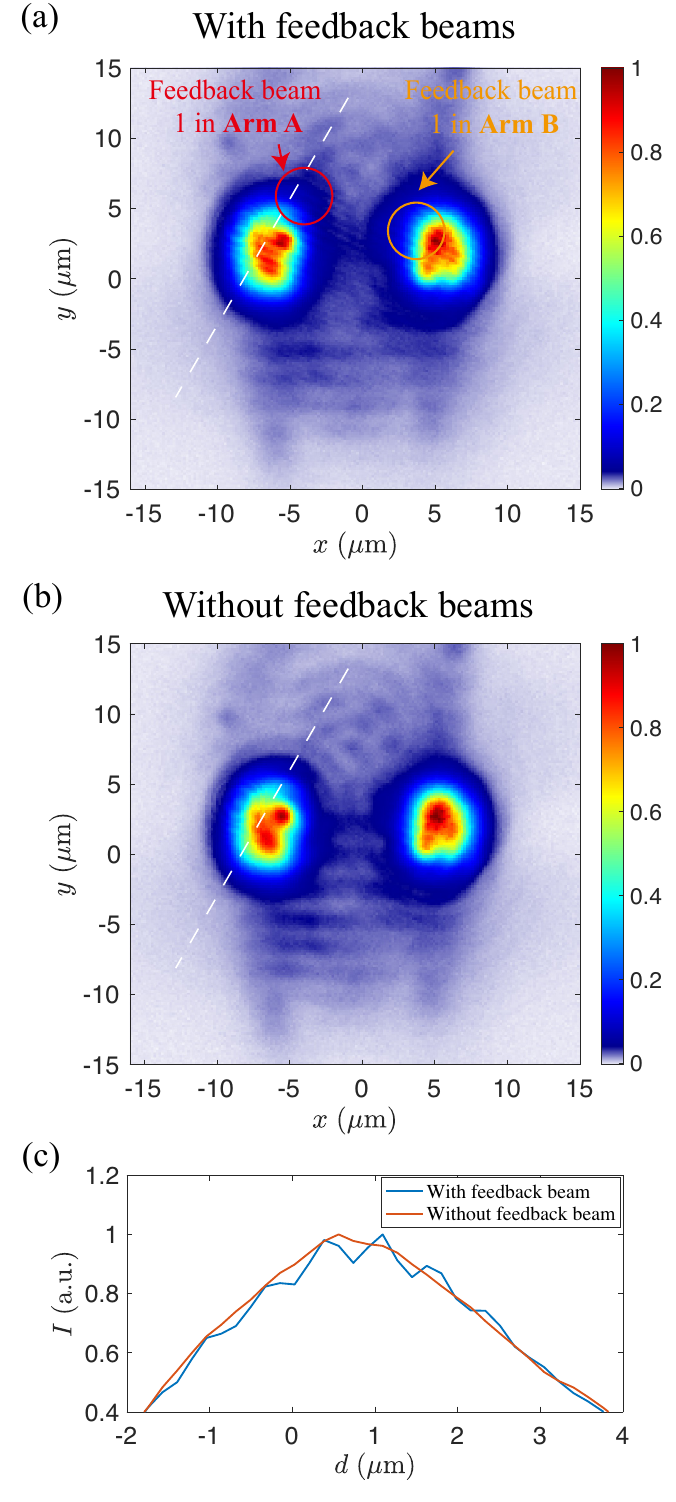}
  \caption{(a) Interference of two condensates through the Michelson interferometer with the feedback light present. (b) The same image under the same experimental conditions, when the feedback light is blocked. (c) The intensity along dashed lines marked in (a) and (b). In (a), the Michelson arm-length difference is set to compensate the external feedback propagation delay $\tau$, so the observed fringes correspond to the first-order cross-coherence evaluated at a time shift $\Delta\simeq\tau$ (not equal-time coherence).
}
  \label{fig5}
\end{figure}

\item {\bf As a control experiment, the feedback light is blocked, and the two condensates no longer are phase locked.}  The interference pattern of the two images occurs {\em only} when the feedback signal is sent from one condensate to the other. Figure~\ref{fig5}(b) shows the interference image under exactly the same conditions as (a), but with the feedback beam to M1 blocked, and Figure~\ref{fig5}(c) shows a comparison of the intensity profile along a cut perpendicular to the fringes for the cases (a) and (b).

\end{enumerate}

\section{Delay-Coupled Exciton--Polariton Condensates: Driven--Dissipative GPE and Phase Dynamics}

Because of the significant length of the external coupling path, the theory of this system must explicitly account for the delay time in the coupling.

We consider two exciton--polariton condensates ($i = 1,2$) confined in 2D harmonic traps and coupled symmetrically via time-delayed feedback. Each condensate is described by a complex wavefunction $\Psi_i(\mathbf{r},t)$ satisfying a driven--dissipative Gross--Pitaevskii equation (GPE):
\begin{eqnarray}
i\hbar\,\frac{\partial \Psi_i(\mathbf{r},t)}{\partial t} &=& 
\biggl[-\frac{\hbar^2}{2m}\nabla^2 + \frac{1}{2}m\hat{\Omega}^2 r^2 + (U_0 - i\hbar \Gamma)|\Psi_i|^2 \nonumber \\
&+& i\hbar P\biggr]\,\Psi_i(\mathbf{r},t)
+ J_{\rm eff}\,e^{i\beta}\,\Psi_j(\mathbf{r-\hat{r}},\,t-\tau), \nonumber\\
\label{main}
\end{eqnarray}
where $m$ is the polariton effective mass,
$\hat{\Omega}$ is the harmonic trap frequency,
 $U_0$ is the polariton--polariton interaction strength,
 $P$ is the effective gain due to incoherent pumping being reduced by the linear losses,
 $\Gamma$ is the nonlinear loss rate,
 $J$ is the coupling strength between condensates (when  coherence filtered, it becomes $J_{\rm eff}= J e^{-\tau/T_{\rm c}}$),
$\tau$ is the round-trip time delay of the feedback path, $T_c$ is the coherence time, and ${\bf \hat{r}}$ is the signal displacement. The static phase offset $\beta$ in the coupling term reflects a global phase shift accumulated during propagation of the leakage photons between the two condensates. In the experimental geometry (mirror-mediated feedback), this phase arises due to the finite optical path length and wavelength and includes dynamical phase from the optical path and the phase shift from reflection. In particular, light leaking from condensate $j$ and reflected by a mirror accumulates a phase $\beta=k L = 2 \pi L/\lambda,$ where 
$L$ is the optical path length from condensate $j$ to $i$ via the mirror,
$ \lambda$  is the photon wavelength (in the cavity),
$k=2 \pi/\lambda$ is the wavenumber of the propagating light.
This phase is static as long as the geometry is fixed but becomes tunable if $L$ is adjusted, e.g. by translating the mirror by a distance $\Delta L$, which shifts $\beta$ by $\Delta\beta = k\Delta L$.
Each mirror reflection can also contribute a constant phase, depending on the Fresnel reflection properties at the mirror surface. 

To elucidate the phase coupling between the condensates, we  make three simplifying assumptions in Eq.(\ref{main}): 
\begin{itemize}
\item (A1) each condensate occupies only the ground harmonic mode
      $\psi_i(\mathbf{r})$ of its trap, so
      $
        \Psi_i(\mathbf{r},t) \simeq \psi_i(\mathbf{r})\,c_i(t),
      $
      with $\psi$ real and  normalized
      $\int |\psi|^2 d^2\mathbf{r}=1$.
\item (A2) Spatial displacement $\hat{\mathbf{r}}$ is neglected on the scale
      of $\psi(\mathbf{r})$ (i.e.\ the coupling samples the same mode).
\item (A3) Pump~$P$ and nonlinear loss~$\Gamma$ balance so that the amplitude
      quickly reaches a steady value; phase evolves on a slower time‑scale.
\end{itemize}

We insert $\Psi_i=\psi_i\,c_i$ into~\eqref{main} and integrate over $\mathbf{r}$.
We define the nonlinear coefficient
$
  U \equiv U_0 \!\int |\psi_i|^4 d^2\mathbf{r},
$
and the single‑particle energy
$
  E_0 \equiv \!\int
      \psi_i^*\!\left(-\frac{\hbar^2}{2m}\nabla^2+\frac12m\hat{\Omega}^2 r^2\right)\!\psi_i
      \,d^2\mathbf{r},
$
which we assume to be equal for the two condensates, since experimentally the two condensates are tuned to have nearly the same energy. This allows us
to obtain two coupled ordinary differential equations
\begin{equation}
i\hbar\,\dot c_i =
\Bigl[E_0 + (U-i\hbar\Gamma)\,|c_i|^2 + i\hbar P\Bigr]\,c_i
\;+\; J_{\rm eff}\,e^{i\beta}\,c_j(t-\tau).
\label{eq:SL}
\end{equation}
We write $c_i(t)=A_i(t)\,e^{i\theta_i(t)}$ with real amplitude
$A_i>0$ and phase $\theta_i$.
Separating real and imaginary parts of~\eqref{eq:SL} gives
\begin{eqnarray}
\hbar\,\dot A_i &=&
 \hbar\bigl(P-\Gamma A_i^{2}\bigr)A_i \label{eq:amp} \\
&& + J_{\rm eff} A_j(t-\tau)\,
   \sin\!\bigl[\theta_j(t-\tau)-\theta_i(t)+\beta\bigr], \nonumber
\\[4pt]
-\hbar A_i\,\dot\theta_i &=&
 \bigl(E_0+U A_i^{2}\bigr)A_i \label{eq:phase} \\
&& + J_{\rm eff} A_j(t-\tau)\,
   \cos\!\bigl[\theta_j(t-\tau)-\theta_i(t)+\beta\bigr].
 \nonumber
\end{eqnarray}

Under assumption (A3) the pump balances loss,
$P-\Gamma A_i^2=0$, so $A_i(t)\to A=\sqrt{P/\Gamma}$.
Amplitude fluctuations relax on the time scale $(2\Gamma A)^{-1}$ and
are neglected henceforth, i.e.\ $A_i(t)\equiv A$.
With $A_i=A_j=A$ fixed, we divide~\eqref{eq:phase} by $A$ to get
$
\dot\theta_i(t)
  \;=\; \omega_0
  \;-\; K_{\rm eff}\,\cos\!\bigl[\theta_j(t-\tau)-\theta_i(t)+\beta\bigr],
$
$K_{\rm eff} \equiv J_{\rm eff}/{\hbar },$
$\omega_0 \equiv -(E_0+UA^{2})/{\hbar}.
$

A cosine coupling can be written as a sine with a phase shift:
$K_{\rm eff}\cos(\Delta)=K_{\rm eff}\sin(\!\tfrac{\pi}{2}-\Delta)$.
Defining
$\alpha\equiv\beta-\frac{\pi}{2}$ we obtain the
Kuramoto--Sakaguchi model with delay
\begin{equation}
\dot\theta_i(t)=\omega_0
  +K_{\rm eff}\,\sin\!\bigl[\theta_j(t-\tau)-\theta_i(t)+\alpha\bigr],
\qquad j\neq i.
\label{eq:KSdelay}
\end{equation}
This equation is invariant under the one-parameter family of rotating-frame transformations for arbitrary $\omega$: $\theta_i\rightarrow \theta_i - \omega t, $ $\omega_0\rightarrow \omega_0 - \omega, $ $\alpha \rightarrow \alpha + \omega \tau.$ Therefore, without loss of generality we set $\alpha=0$ for $\tau\ne 0$.

In experiments, both condensates can be detuned from $\omega_0$ so we consider a more general case of  a  two–oscillator system
\begin{equation}
\begin{aligned}
\dot{\theta}_{1}(t)&=\omega_{1}
        +K_{\rm eff}\sin\!\bigl[\theta_{2}(t-\tau)-\theta_{1}(t)\bigr],\\
\dot{\theta}_{2}(t)&=\omega_{2}
        +K_{\rm eff}\sin\!\bigl[\theta_{1}(t-\tau)-\theta_{2}(t)\bigr].
\end{aligned}\label{eq:twoosc-alpha}
\end{equation}
 and introduce the following notation:
\begin{eqnarray}
\Phi=\frac{\theta_1+\theta_2}{2},\quad
\phi=\frac{\theta_2-\theta_1}{2},\quad
\\
\bar\omega=\frac{\omega_1+\omega_2}{2},\;
\Delta\omega=\frac{\omega_2-\omega_1}{2}.
\end{eqnarray}
Denoting $\widehat{(\,\cdot\,)}:= (\cdot)(t-\tau)$, elementary
sum– and difference–angle identities give the exact delay system
\begin{subequations}\label{eq:Phi-phi}
\begin{align}
\dot\Phi &= \bar\omega
  +K_{\rm eff}\;\sin(\widehat{\Phi}-\Phi)\;
         \cos(\widehat{\phi}+\phi),\label{eq:Phi}\\[4pt]
\dot\phi &= \Delta\omega
  -K_{\rm eff}\;\cos(\widehat{\Phi}-\Phi)\;
         \sin(\widehat{\phi}+\phi).\label{eq:phi}
\end{align}
\end{subequations}
%--------------------------------------------------------------

For a uniformly rotating solution  
$\Phi(t)=\Omega t,$ with the constant phase difference $\phi(t)=\phi^{*}$, these equations reduce to
\begin{align}
    \Omega \;=\; \bar\omega
                \;-\; K_{\rm eff}\cos2\phi^{*}\,
                       \sin(\Omega\tau), \label{A}\\
    0 \;=\; \Delta\omega
          \;-\; K_{\rm eff}\cos(\Omega\tau)\,
                 \sin 2\phi^{*}.
                 \label{B}
\end{align}

\medskip
If the oscillators are identical ($\Delta \omega=0$), then, unless $\cos(\Omega \tau)=0$ exactly, we have $\sin 2\phi^{*}=0$, hence  
  $\phi^{*}=0$ (in‑phase) or $\phi^{*}=\pi/2$ (anti‑phase).  
  Substituting each value into Eq.~(\ref{A}) gives
$
  \Omega=\bar\omega\pm K_{\rm eff}\sin(\Omega\tau),
  $
  which matches Yeung and Strogatz Eq.\,(5) \cite{yeung1999time}.

\begin{figure}[t]
  \centering
  \includegraphics[width=.9\columnwidth]{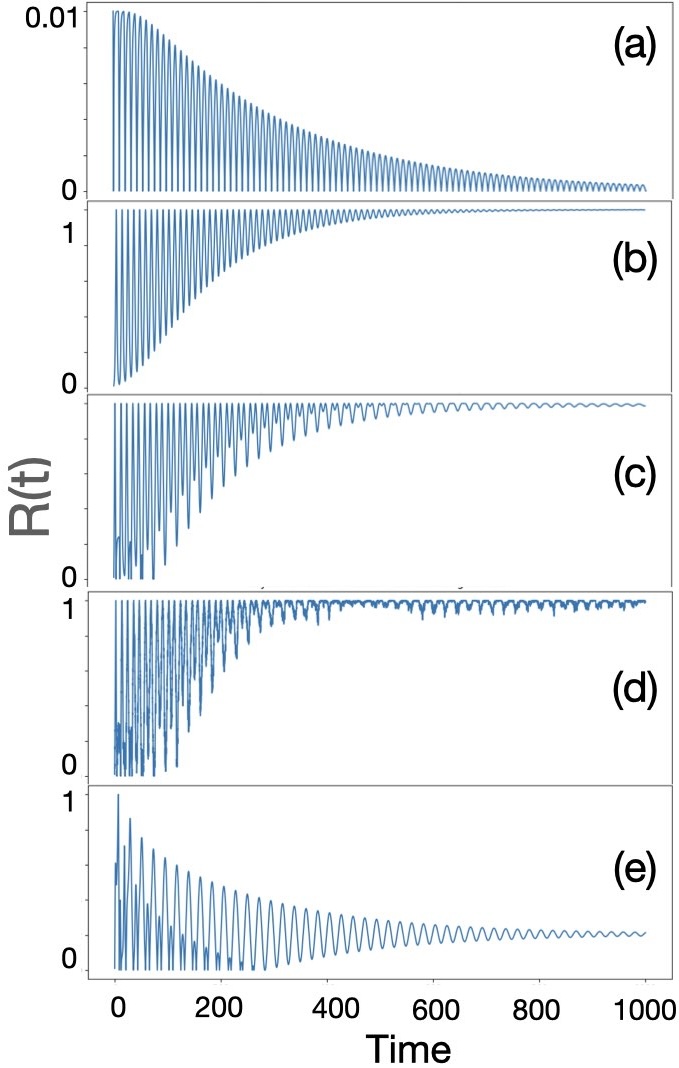}
  \caption{Temporal evolution of the order parameter
$
R(t)=|e^{i\theta_1(t)}+e^{i\theta_2(t)}|/2
$
for two delay‐coupled oscillators governed by Eqs.~(\ref{noise}), characterized by natural frequency $\omega_0$, detuning $\Delta\omega$, coupling strength $K$, time delay $\tau$, and noise amplitude $\sigma$. The phase history on the interval $t\in[-\tau,0]$ is prescribed as
$
\theta_1(t)=\theta_{\mathrm{ic}},$ and $
\theta_2(t)=\pi-\theta_{\mathrm{ic}},$ with $
\theta_{\mathrm{ic}}=10^{-2}.
$ Simulation parameters are
 (a) $  \omega_0=0, K_{\rm eff}=-1,  \Delta\omega=0, \sigma=0;$ (b) $\omega_0=0, K_{\rm eff}=1, \Delta\omega=0, \sigma=0;$  (c) $\omega_0=0, K_{\rm eff}=1, \Delta\omega=0.25, \sigma=0;$ (d) $\omega_0=0, K_{\rm eff}=1, \Delta\omega=0.25, \sigma=0.1;$ (e) $  \omega_0=1, K_{\rm eff}=1,  \Delta\omega=0.4, \sigma=0.$}
  \label{fig:R}
\end{figure}

In the system with detuning  $\Delta \omega\ne 0$,
  $\sin 2\phi^{*}\neq0$ in general, so the
  $\cos(2\phi^{*})$ term in Eq.(\ref{B}) must be retained.
  Solutions with any $2\phi^{*}\neq 0,\pi$ (mod $2\pi$) are possible provided
  $|\Delta\omega|\le| K_{\rm eff}\cos(\Omega\tau)|$, and the exact
  offset is obtained by solving the coupled transcendental
  system Eqs.(\ref{A}-\ref{B}).
We conclude that delay alone does not create a stable non‑trivial phase difference between
identical oscillators; the generic steady offsets are $0$ or $\pi$.
Detuning, noise (effective detuning) or an explicit frustration phase
$\alpha$ are required to obtain a locked offset distinct from those two
values. Figure \ref{fig:R} depicts the solutions of the coupled oscillators model with detuning and noise
\begin{equation}
\begin{aligned}
d\theta_1(t)
&=  (\omega_0+\Delta\omega)\,dt
  + K_{\rm eff}\,\sin\bigl(\widehat{\theta_2}-\theta_1\bigr)\,dt
  + \sigma\,dW_1(t),\\
d\theta_2(t)
&= (\omega_0-\Delta\omega)\,dt
  + K_{\rm eff}\,\sin\bigl(\widehat{\theta_1}-\theta_2\bigr)\,dt
  + \sigma\,dW_2(t),
  \label{noise}
\end{aligned}
\end{equation}
where $W_1,W_2$ are independent Wiener processes and are discretized as
$\sqrt{dt}\,\xi_i$, $\xi_i\sim\mathcal N(0,1)$.
We plot the order parameter $R(t)=|e^{i \theta_i} + e^{i \theta_2}|/2$ introduced by Kuramoto \cite{kuramoto1984chemical}. $R(t)$ measures the system’s phase coherence. In
particular, $R=1$ if the oscillators are in-phase,
whereas $R(t)=0$ if the oscillators have $\pi$ phase difference. Figure \ref{fig:R}(e) depicts the convergence of the system to the phase difference $\phi^*\approx 1.356$ different from both 0 and $\pi$ ($R(t)\approx 0.213$).

For $N$ condensates with pairwise couplings $J_{ij}$, phase lags
$\beta_{ij}$, self-frequencies $\omega_i,$ and  distinct delays $\tau_{ij}$,
the phase dynamics become
\begin{equation}
\dot\theta_i(t)
   \;=\;\omega_i
   +\sum_{j\neq i} K_{ij}^{\rm eff}\,
     \sin\!\bigl[\theta_j(t-\tau_{ij})-\theta_i(t)+\alpha_{ij}\bigr],
     \label{eq:Nnode}
\end{equation}
with $K_{ij}^{\rm eff}=J_{ij} e^{-\tau_{ij}/T_{\rm c}}/{\hbar},$ 
$\alpha_{ij}=\beta_{ij}-\tfrac{\pi}{2}.$
Yeung and Strogatz showed that introducing a finite delay $\tau$ in the system of Kuramoto oscillators produces qualitatively new phenomena \cite{yeung1999time}. Notably, multiple synchronized states can exist (with different collective frequencies or phase-lags), and a stable incoherent state can coexist with stable synchrony (bistability). Additionally, one can observe unsteady oscillatory dynamics of the order parameter $\sum e^{i \theta_i}/N$ in certain parameter regimes, in contrast to the monotonic approach to steady state in the classic Kuramoto model.

%=====================================================================
\section{Why phase coherence survives a $\tau\!\sim\!1.6\,$~ns delay}

Two exciton–polariton condensates separated by a $L\!\simeq\!24\,$cm
external cavity remain phase locked and display clearly visible
interference fringes even though the single‑trap first‑order coherence
decays after only
$
\sim141 ~\mathrm{ps} .
%<T_{\mathrm c}< 1~\mathrm{ns}.
$

The interferometric measurement is performed with a compensating delay, so it probes a time-lagged first-order correlation rather than equal-time coherence at the sample.
In the delay-locked regime the relevant phase relation is between $\theta_2(t)$ and $\theta_1(t-\tau)$, not between $\theta_2(t)$ and $\theta_1(t)$ at the same time.
Equivalently, the first-order cross-coherence is expected to peak near $\Delta=\tau$:
for an interferometer delay $\Delta=\tau+\delta$, the fringe visibility should decrease as $|\delta|$ exceeds the intrinsic coherence time scale $T_c$.
Thus the observation of fringes at $\Delta\simeq\tau$ should be interpreted as delay synchronization (time-lag locking), not as an intrinsic coherence time of order $\tau$.

\begin{figure}[t]
  \centering
  \includegraphics[width=.9\columnwidth]{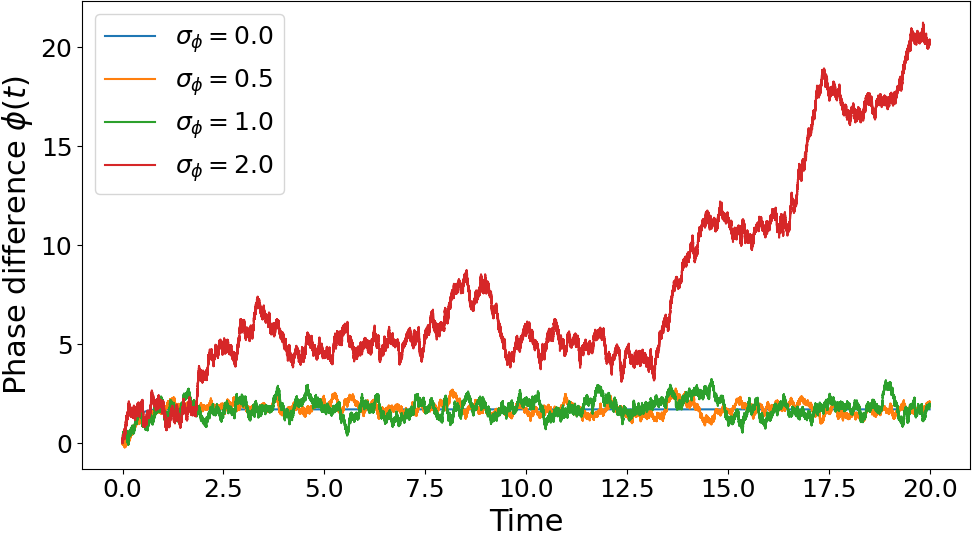}
  \caption{Temporal evolution of the phase difference 
$
\phi(t)=(\theta_2-\theta_1)/2
$
for two delay‐coupled oscillators governed by the noisy Adler equation Eqs.~(\ref{adler}) starting from the same initial condition $\phi(0)=0$ for different levels of noise.  The simulation parameters are $K=100, \tau=1.6, T_c=0.5,$ $\Omega=10$, $\Delta\omega=1$. The fixed point is $\phi^*\approx 1.7$, so that $\sin(2 \phi^*)\approx -0.256 = \Delta\omega/K_{\rm eff} \cos(\Omega \tau)$ as expected. For clarity of the drift the phase difference was not wrapped around $2\pi$. The parameters are in normalized units.
}
  \label{fig:Adler}
\end{figure}

The round‑trip delay is
$
\tau=2L/c
     \simeq 0.48\text{ m}/3\times10^{8}\text{ m/s}
     \simeq 1.6\;\mathrm{ns},
$
 hence $e^{-\tau/T_{\rm c}} \sim 10^{-5}
 $.  
At first sight this seems to contradict the simple estimate that 
this  would suppress the
effective coupling by a factor $
\lesssim 10^{-5}$ for the shortest $T_{\rm c}$.  The phase locking can be explained by noting that 
the mirror feedback acts as a classical injection
      locking term.  
      Both traps receive the same delayed field, so the stochastic
      contribution can be decomposed into an identical part
      $\xi_{\mathrm c}(t)$ and two independent parts
      $\xi_{1,2}(t)$.  
      Only the difference $(\xi_2-\xi_1)$ enters the
      phase‑difference dynamics, reducing the effective diffusion to
      $\sigma_{\phi}\ll \sigma$.
We rewrite Eqs.(\ref{noise}) to incorporate these effects 
 as
\begin{align}
\dot\theta_1 &= \omega_1
      +\xi_{\mathrm c}(t)+\xi_1(t)
      +K\,e^{-\tau/T_{\mathrm c}}
        \sin\!\bigl[\theta_2(t-\tau)-\theta_1(t)\bigr], \notag\\[4pt]
\dot\theta_2 &= \omega_2
      +\xi_{\mathrm c}(t)+\xi_2(t)
      +K\,e^{-\tau/T_{\mathrm c}}
        \sin\!\bigl[\theta_1(t-\tau)-\theta_2(t)\bigr],         \label{eq:modNoise}
\end{align}
with noise correlators
$
   \langle\xi_{i}(t)\xi_{j}(t')\rangle
      =2\sigma_{\text{ind}}\delta_{ij}\delta(t-t'),
\;
   \langle\xi_{\mathrm c}(t)\xi_{\mathrm c}(t')\rangle
      =2\sigma_{\text{com}}\delta(t-t').
$
Here, $\sigma$ is the  total phase–diffusion rate (linewidth) of one isolated
   condensate ($\sigma=1/T_{\mathrm c}$); $\sigma_{\rm ind}$ is the diffusion rate of the independent noise component that remains
   uncorrelated between the two traps; $\sigma_{\rm com}$ is the diffusion rate of the common‑mode noise component delivered
   through the shared mirror path; $\sigma_\phi$ is the diffusion of the relative phase
   $\phi$.
For the relative phase
$\phi$ this yields
\begin{equation}
\dot\phi = \Delta\omega
        -K\,e^{-\tau/T_{\mathrm c}}
          \cos(\Omega\tau)\sin 2\phi
        +\xi_\phi(t),
        \label{adler}
\end{equation}
where we included the independent stochastic torque
$\xi_{\phi}(t)$ with diffusion $\sigma_{\phi}$ so that $\langle\xi_{\phi}(t)\xi_{\phi}(t')\rangle=2\sigma_{\phi}\,\delta(t-t')$.  Equation (\ref{adler}) is a noisy Adler
equation \cite{adler1946study}.
Locking occurs when
\begin{equation}
   |\Delta\omega|
         \;\le\;
   e^{-\tau/T_{\mathrm c}}\,
   |K\cos(\Omega\tau)|.
   \label{eq:criterion}
\end{equation}
Without the noise, Eq. (\ref{adler}) possesses a stable fixed point
$
\phi^{*}
$
such that $\sin(2 \phi^*)=\Delta \omega/K_{\rm eff}\cos(\Omega\tau).$
Linearizing about $\phi^{*}$ yields an Ornstein–Uhlenbeck process \cite{uhlenbeck1930theory} with
effective damping $K_{\rm eff}\cos(\Omega\tau)\cos\phi^{*}$, hence a narrowed
linewidth
\begin{equation}
\gamma_{\text{locked}}
   \;\approx\;
   \frac{\sigma_{\phi}}{|K_{\rm eff}\cos\phi^{*}|}
   \;\ll\;
   \sigma.
\label{A4}
\end{equation}
We illustrate the behaviour of the system by numerically integrating Eq.~(\ref{adler}) for suitable system parameters and various levels of noise in Fig. ~(\ref{fig:Adler}). The system parameters satisfy Eq.~(\ref{eq:criterion}) for the existence of fixed point Eq.~(\ref{eq:criterion}). For small to intermediate noise the phase difference drifts around the fixed point of Eq.~(\ref{adler}). For strong noise the trajectory is no longer fluctuates around that value and the coherence is lost.

We conclude that strong mutual injection converts two
broad, uncorrelated sources into a single collective oscillator with a
linewidth orders of magnitude narrower than $\sigma$.
The factor  $e^{-\tau/T_{\mathrm c}}$ is small
for the present device parameters but $\sigma_{\phi}\ll K_{\rm eff}$, so the mirror‑coupled
condensates may still satisfy the inequalities for phase locking, explaining the robust
fringe visibility seen in the experiment even at round‑trip distances
of $\sim\!24$cm.

\section{CONCLUSIONS AND OUTLOOK}

We have demonstrated phase locking of two condensates via external feedback where there is no measurable coupling between the condensates due to their proximity to each other. Although we have used only two condensates, and the interference fringe visibility is only about 10\%, this is an important proof-of-principle that opens up a fascinating new possibility for polariton condensates in lattices. Namely, unlike any other lattice system we know of, this system allows coupling between lattice sites at a distance, instead of just nearest neighbors or further neighbors at decreasing strength. This opens up the possibility of programming arbitrary coupling between lattice sites using the external optical system.

As discussed at length in the Supplemental Material, the experimental realization of tunable, time-delayed, mirror-mediated coupling between exciton–polariton condensates  provides a powerful hardware platform for several machine learning (ML) techniques. Polariton systems with delayed, nonlocal coupling uniquely meet key neuromorphic requirements such as programmability of connections, ultrafast speed, and broad bandwidth, enabling both efficient inference and on-chip training.

This means that learning can occur within the physical device where weights can be adjusted either optically or via internal self-organization while closing the loop for a fully trainable neuromorphic platform. In summary, exciton–polariton condensate hardware is uniquely poised to realize physically trainable neural networks, where both forward inference and feedback-based learning are implemented by the same optical processes. The mirror-mediated coupling introduced here is the key enabler: it provides reconfigurable, ultrafast links that can broadcast phase information and error signals across the network with minimal latency, fulfilling the essential requirements of programmability, speed, and bandwidth for next-generation ML hardware.

%%%%%%%%%%%%%%%%%%%%%%%%%%%%%%%%%%%%%%%%%
\section{Acknowledgements}
This project has been supported by the National Science Foundation grant  DMR-2306977. N.G.B  acknowledges the support from HORIZON EIC-2022-PATHFINDERCHALLENGES-01 HEISINGBERG Project 101114978,  Weizmann-UK Make Connection Grant 142568, and  the EPSRC UK Multidisciplinary Centre for Neuromorphic Computing award UKRI982.

\section{Disclosures}
The authors declare no conflicts of interest.

\section{Data Availability Statement}
Data underlying the results presented in this paper are not publicly available at this time but may be obtained from the authors upon reasonable request.

\clearpage
\bibliography{references.bib, nb-refs}

\clearpage
\title{Supplementary Materials for: Mirror-mediated long-range coupling and robust phase locking of spatially separated exciton-polariton condensates}
\date{\today}
\maketitle
\beginsupplement

\section{Design of the sample}
As mentioned in the main text, the microcavity sample consisted of a total of 12 GaAs quantum wells with AlAs barriers embedded within a distributed Bragg reflector (DBR). The quantum wells are in groups of four, with each group placed at one of the antinodes of the $3\lambda/2$ cavity. The DBRs are made of alternating layers of AlAs and \ce{Al_{0.2}Ga_{0.8}As}. The top DBR is composed of 23 pairs and the bottom DBR is composed of 40 pairs.

\section{Laser beam in real space}
\begin{figure}
  \centering
\includegraphics[width=0.45\textwidth]{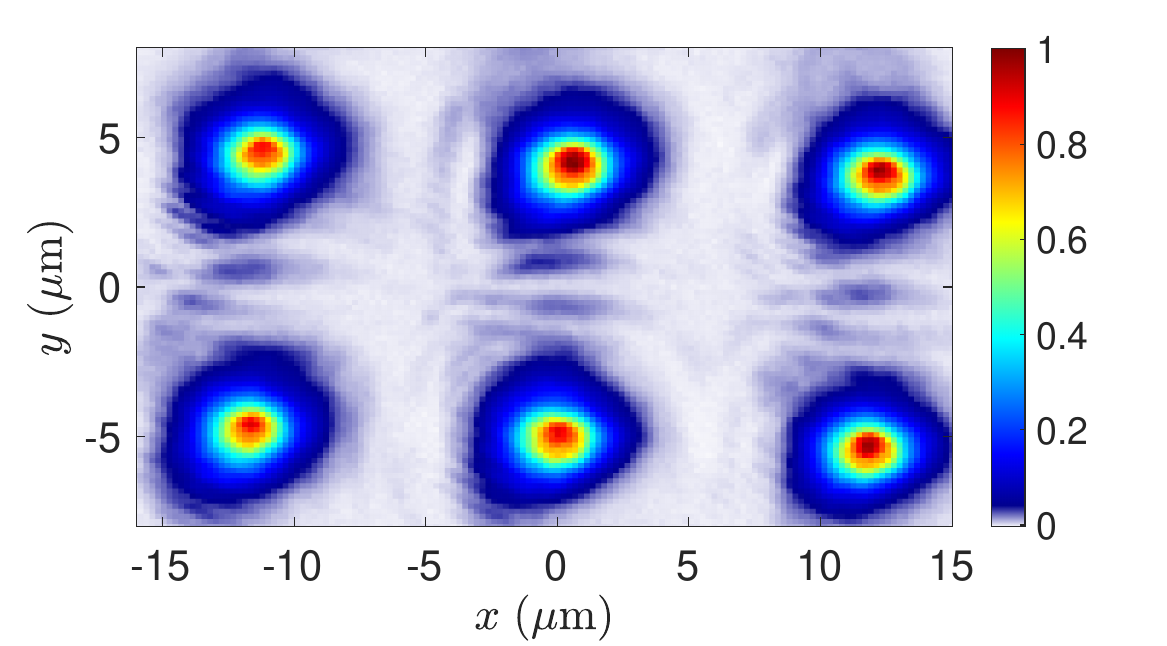}
  \caption{Real-space image of laser beam generated by using SLM. }
  \label{figS1}
\end{figure}

We used an spatial light modulator (SLM) to reshape the laser to a pattern of six spots, as shown in Figure~\ref{figS1}, which enabled the formation of two traps and the realization of polariton condensates. Each of the six excitation spots functions both as a polariton source and as an energy barrier, forming a trapping potential between each four spots.
\section{Self-Coherence of the Condensate and Comparison to Phase-Locked Visibility}
\begin{figure}
  \centering
  \includegraphics[width=1\columnwidth]{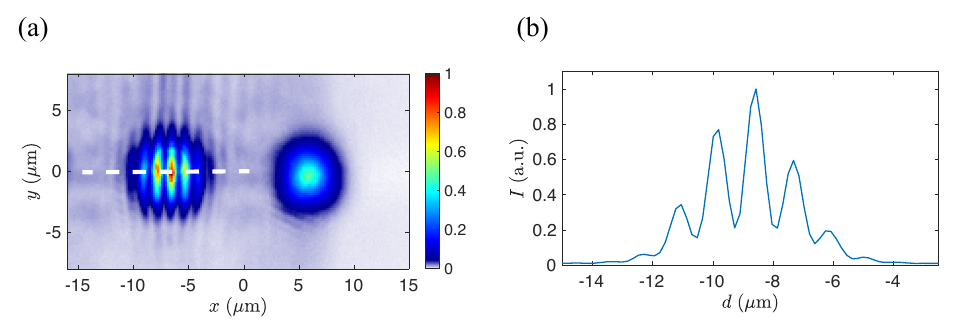}
  \caption{Self-coherence of the condensate. (a) Interference pattern of the condensate with itself. (b) Fringe visibility extracted along the white dashed line in (a).}
  \label{figR1}
\end{figure}

To provide a quantitative measure of the phase-locking effect, we compare the fringe visibility in two cases:
(1) the self-coherence at zero delay time, as shown in Fig.~\ref{figR1}(a) and (b), where the visibility is 65\%; and
(2) the coherence between two condensates after phase locking, as shown in Fig.~5(c) in the main manuscript, where the visibility is approximately 10\%.
The observation of a finite, nonzero fringe visibility in the phase-locked configuration suggests the presence of synchronization between the two condensates. The reduced visibility compared to the self-coherence case indicates that the relative phase is not fully stabilized and is subject to residual fluctuations. This comparison therefore provides a useful, experimentally accessible measure to characterize the presence and relative strength of synchronization in our system.

\begin{figure}
  \centering \includegraphics[width=1\columnwidth]{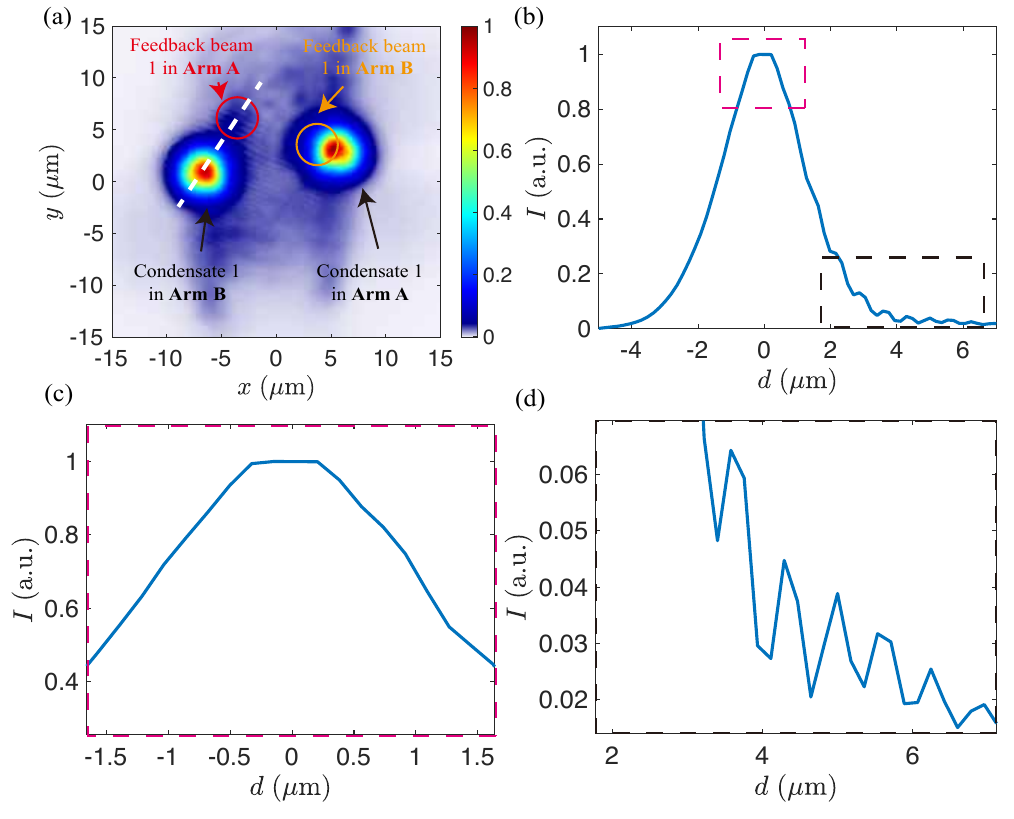}
      \centering
      \caption{Interference between the image of Condensate 1 and its feedback beam through Arm A. (a) Real-space image of the condensate and its feedback beam. (b) Fringe visibility extracted along the white dashed line in (a), with the red and black regions corresponding to the locations of the condensate and the feedback beam, respectively. (c),(d) Zoomed-in views of (b).}
  \label{figR2}
\end{figure}

We address the possibility of coherence between the feedback beam originating from condensate 1 and condensate 1 itself. To ensure that such an effect does not influence our conclusions, we provide additional analysis of the fringe visibility in Fig. 4 of the main manuscript, as shown in Fig.~\ref{figR2}(a).

The fringe visibility is extracted along the white dashed line and plotted in Fig.\ref{figR2}(b). We compare the visibility in two spatial regions: the red region corresponds to the condensate itself, while the black region corresponds to the coherence between the feedback beam and the condensate. Zoomed-in views of these two regions are shown in Fig.\ref{figR2}(c) and (d), respectively.

Due to the intentional spatial offset between the condensate and the feedback beam, interference fringes are observed only at the location of the feedback beam. As a result, the coherence between the feedback beam and the original condensate does not affect the interference signal analyzed in Fig. 5 of the main manuscript.\\

\section{DEFINING threshold POWER}
\begin{figure}
  \centering  \includegraphics[width=0.8\columnwidth]{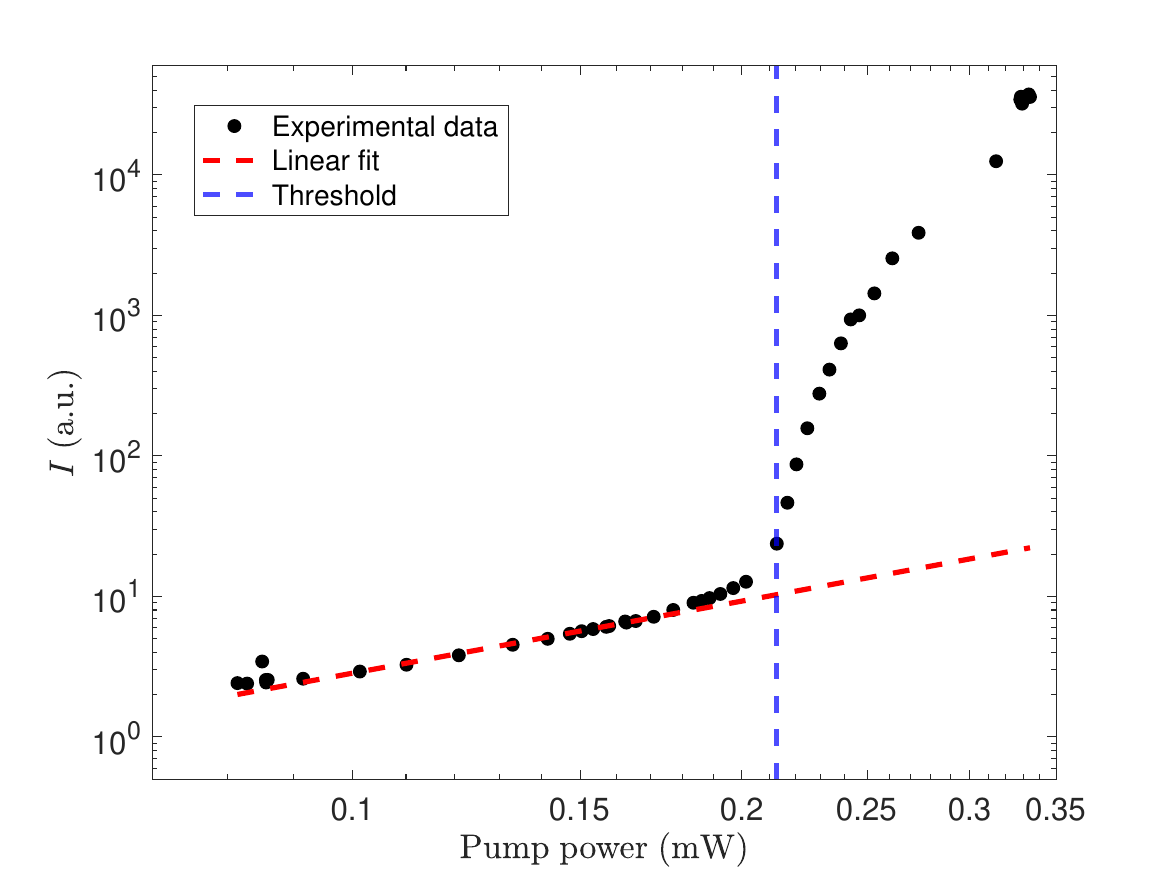}
  \caption{Threshold power of condensate. The blue dashed line indicates the condensate threshold power, corresponding to approximately 0.2 mW. }
  \label{figS2}
\end{figure}

To determine the threshold power of the BEC, we extracted the intensity of the polaritons at $k=0$ from the angle-resolved PL and plotted it as a function of pump power as shown in Fig. \ref{figS2} to obtain the "S" curve. To insure that we are only collecting PL from the polaritons inside the trap, we used a pinhole in real-space to collect PL from only one condensate.
\par
Near the condensation threshold, a nonlinear increase in intensity is observed, which becomes linear again at much higher pump power. We define the threshold power when the measured "S" curve deviates from being linear by approximately $12\%$. The extracted threshold power is approximately 0.2 mW.

\section{Coherence time and phase locking delay}
\begin{figure}
  \centering \includegraphics[width=0.9\columnwidth]{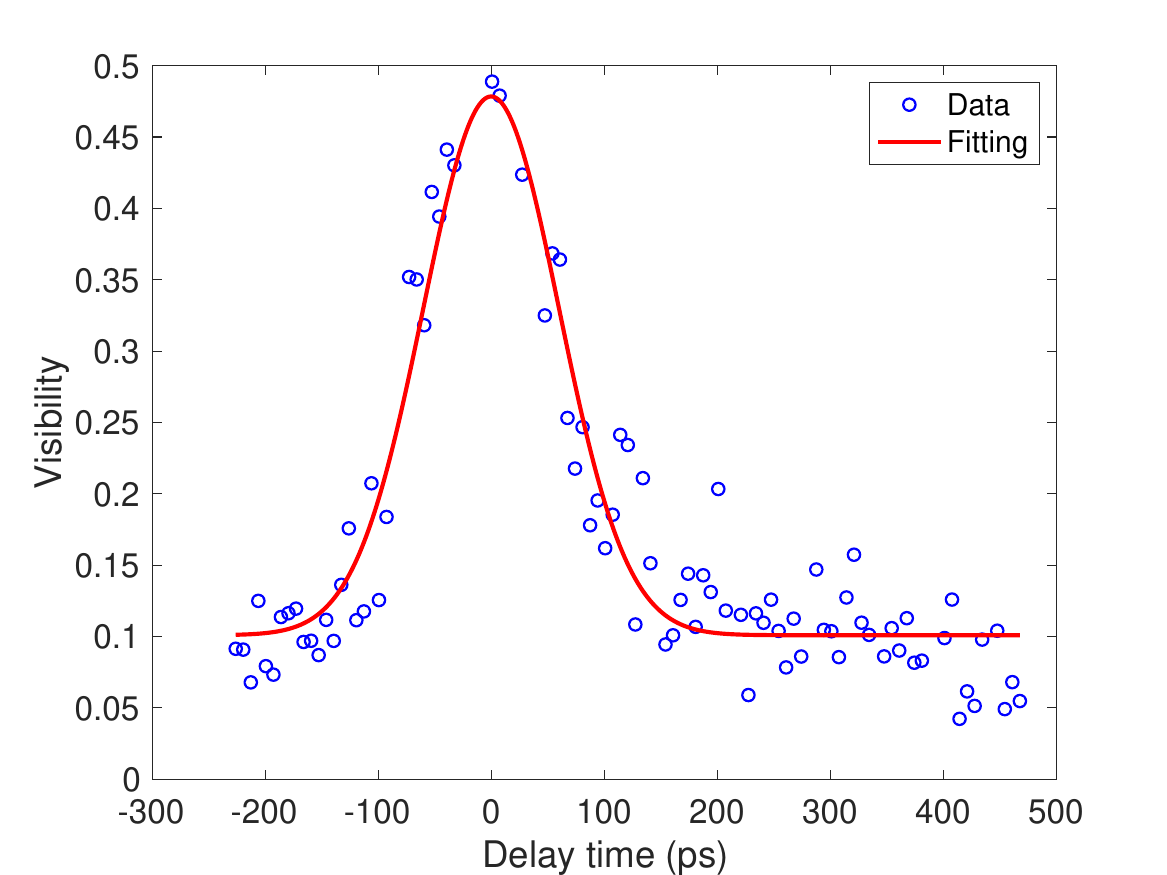}
      \centering
      \caption{Coherence time of the condensate. Circles: the measured visibility as a function of time-delay. Solid line: a fit to the data using a Gaussian function plus a constant offset. From the best fit, we obtain a FWHM of 141 ps.}
  \label{figR3}
\end{figure}

To measure the coherence time, we send a single condensate into a Michelson interferometer. By varying the delay time between the two arms, we plot the fringe visibility as a function of delay time in Fig.~\ref{figR3}, which is well fitted by the Gaussian curve shown in red. Taking the full width of the Gaussian shape, the coherence time of the condensate is estimated to be approximately 141 ps. 

The coherence time of the condensate is much shorter than the phase delay; nevertheless, we can still observe phase locking. This is because, in the phase-locking measurement, we introduce a time delay in one arm of the Michelson interferometer that matches the optical path length of the feedback path.

To understand this, consider a condensate (denoted A) with a phase at a given time $\theta_A(t)$ that fluctuations on a timescale set by the coherence time $\tau$. If the PL of condensate A is fed back to condensate B after propagating over a distance $d$, corresponding to a delay time $t_d$, then condensate B locks to the phase of a condensate A at an earlier time $\theta_A(t-t_d)$. As a result, if now in the Michelson interferometer, a delay is introduced in one of the arms to interfere condensate B with condensate A, with a delay time $t_d$, then fringes would be observed even when $t_d$ is larger than the coherence time of condensate A. \\
\par
This is supported by the observation in the main text that when a condensate is interfered with its own feedback, interference fringes are still observed even when the feedback path length corresponds to a delay much longer than the intrinsic coherence time. Crucially, the fringes appear only when one arm of the Michelson interferometer is delayed such that the condensate emitted at time \(t_0\) interferes with its own feedback at time \(t - t_0\).

\section{Condensate 2 }

In the main text, we primarily labeled Condensate 1 and its feedback beam. However, the feedback beam from Condensate 2 is also present within the field of view, and we will describe this in this section.

\begin{figure}[htbp]
  \centering
\includegraphics[width=0.45\textwidth]{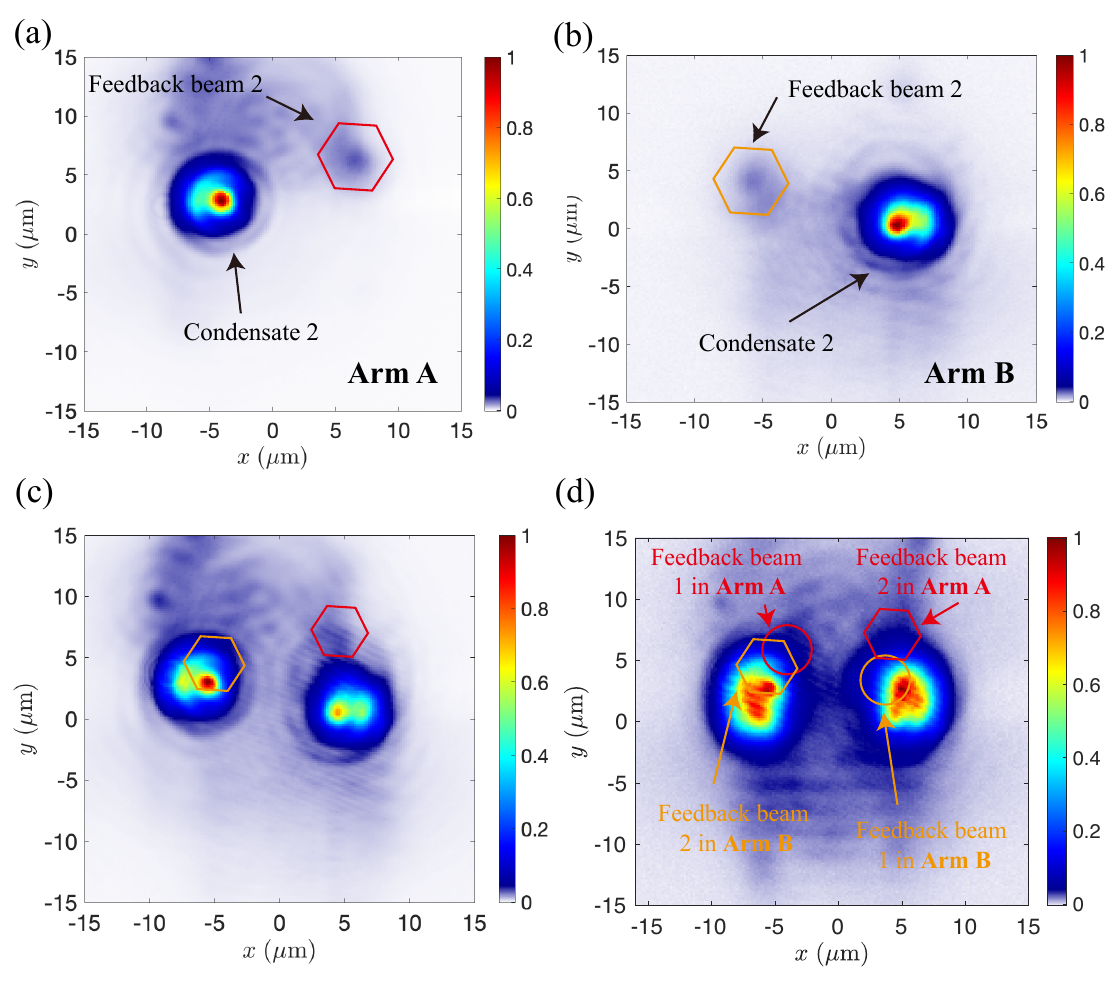}
  \caption{(a),(b) Real-space images of Condensate 2 and its feedback beam in the Michelson interferometer. (c) Interference of the image of Condensate 2 and its own feedback beam through Arm A, with the same image
reversed, through Arm B. (d) Interference of two condensates through the Michelson interferometer with the feedback light present. }
  \label{figS5}
\end{figure}

 As shown in Fig.~\ref{figS5}, when only Condensate 2 is generated, the positions of feedback beam 2 in the two arms of the Michelson interferometer are indicated in Fig.~\ref{figS5}(a) and (b). We also observe clear interference fringes between Condensate 2 and its own feedback beam, as marked by the red indication in Fig.~\ref{figS5}(c). Here, the real-space image of Condensate 2 differs slightly from that shown in Fig. 3(b) of the main text due to small experimental vibrations. However, this does not affect the spectral measurements or any other observations.

In Fig.~\ref{figS5}(d), all feedback beams originating from the two interferometer arms are labeled. In Fig.5(a) of the main text, we only highlighted the interference fringes on the left side to demonstrate phase locking. However, clear interference fringes are also visible in the spot pattern on the right side, indicating that the phase locking is mutual between the two condensates.

\section{Applications to Machine Learning}

In this section we review the potential applications of the method of non-nearest-neighbor coupling in lattices.  We outline three core application areas below.

{\bf Analog Optimization via XY/Kuramoto Energy Landscapes.} Exciton–polariton condensate networks can function as analog spin systems that naturally minimize XY or Kuramoto-model “energy” functions. Each condensate’s phase plays the role of a classical spin, and programmable mirror-couplings act as tunable spin–spin interactions. The polariton network thus evolves toward phase configurations that extremize a given cost function, effectively solving optimization problems in hardware.  
In fact, polariton simulators have demonstrated the ground-state computation of the classical XY Hamiltonian \cite{Berloff2017}, and even “exotic” regimes like cluster synchronization and spin-glass phases can emerge under appropriate coupling patterns \cite{kalinin2018matter, kalinin2019polaritonic}. 

Such analog optimizers are similar to optical Ising machines, but here the continuous phase variables and time-delay elements allow exploring a rich landscape of minima (and potentially escaping local traps via oscillatory dynamics). The mirror-mediated coupling is crucial since  it offers fine programmability of interaction strength and delay, letting one encode arbitrary problem graphs in the condensate network. Moreover, because polariton interactions occur on picosecond scales, the optimization settles orders of magnitude faster than electronic or digital approaches, illustrating the speed advantage of this hardware.

{\bf Reservoir Computing via Delay-Induced Fading Memory.} Time-delayed coupling endows the polariton system with a built-in short-term memory, which is a hallmark of reservoir computing. In a delay-coupled oscillator, past states (within the delay window $\tau$) continue to influence the present, allowing it to encode temporal patterns in a high-dimensional trajectory. This high dimensionality, combined with the nonlinear response of polariton condensates, enables complex time-dependent data processing (e.g. speech or signal recognition) without needing explicit training of internal weights.
 The fading memory means the system naturally “remembers” recent input history over timescales set by $\tau$, then gradually forgets. This is a desirable property for tasks like temporal pattern classification and forecasting. 
 
 Delay-coupled electronic and photonic reservoirs have already achieved tasks such as spoken digit recognition and time-series prediction, and the polariton platform offers the same capability at far higher speeds and lower energy. In our polariton implementation, a mirror is used to feed back the condensate’s emission with a controllable delay, effectively creating a loop memory. By injecting information as modulated optical pulses, one can drive the condensate network into a unique transient response that encodes the input sequence in its phase and amplitude dynamics. Because the condensates are nonlinear oscillators, different inputs generate separable trajectories in phase space, which a simple readout can classify. Recent work indeed demonstrated a time-delay polariton reservoir performing an XOR timing task: two polariton condensates were coupled via a delayed optical feedback, and their ps-scale oscillatory response successfully encoded the logic needed for XOR processing
\cite{mirek2022neural}. There the long-lived excitonic reservoir in the microcavity  provides a memory buffer (via slowly decaying excitons) that, in combination with the fast photon feedback, creates a multi-timescale system.

{\bf Neuromorphic Inference and On-Chip Learning with Phase-Based Dynamics}. Beyond reservoir-style processing, exciton–polariton hardware can implement structured neural network architectures for pattern recognition and inference. By using phase-coded signals and wave propagation through the condensate lattice, one can realize both feed-forward and recurrent neuromorphic networks.  

Entire feed-forward networks have been built from such polariton “neurons”  using spatial light modulators to set coupling weights  and achieved competitive accuracy on benchmarks like MNIST digit classification \cite{opala2019neuromorphic}.
 Notably, due to the strong polaritonic nonlinearity, these optical networks operate with extraordinary energy efficiency: a single synaptic operation was realized with $~16$ pJ of optical energy
on par with the best electronic neuromorphic chips, while exploiting massive parallelism and light-speed signal propagation. 
 
 Importantly, this polariton platform supports not only neural inference but also physical training through on-chip learning rules. The mirror-mediated symmetric couplings and continuous dynamics satisfy the conditions for physical back-propagation algorithms such as Equilibrium Propagation (EP) and Contrastive Hebbian Learning (CHL). In these schemes, the network itself computes weight updates by responding to slight perturbations, eliminating the need for external gradient calculations. 
 
 For instance, it was recently demonstrated that a network of coupled phase oscillators (governed by the Kuramoto/XY model, as in our polariton system) can be trained via EP \cite{wang2024training}.
In EP, one first lets the physical network settle to an equilibrium with a given input (free phase), then “nudges” the output nodes toward a target pattern and allows the system to relax to a new equilibrium (perturbed phase). The locally measured changes in each connection during this process directly indicate the weight gradient.

Because exciton–polariton condensates obey tunable nonlinear dynamics and have optically accessible states, one can implement this procedure in hardware – the gradients are obtained by purely local optical measurements of phase/intensity shifts, rather than by digital backpropagation. Likewise, CHL provides a framework for weight update by running the network in two phases (one with outputs clamped to desired values, one free-running) and applying a Hebbian update based on the difference in correlations
This was  proposed  in the context of Hopfield-like analog networks \cite{baldi1991contrastive}
 and it maps naturally onto the polariton system: one can impose a target output by injecting an appropriate optical field into the output condensate(s) (clamped phase), then remove it (free phase), and the mirror-coupled interactions will automatically propagate the error signals. The symmetric mirror coupling ensures that feedback and feedforward paths are equivalent (a requirement for CHL/EP convergence
), something difficult to achieve in conventional electronics but inherent to our polaritonic design. 

{\bf Speed, Parallelism and Hierarchical Delays.} 
Even though the longest feedback loop in our present setup is of the order of the round-trip time of approximately \(1\)ns (mirror separation $\sim24$cm), each condensate will update its internal phase on the picosecond timescale set by the photon lifetime $(\tau_{\rm rad}\approx 50$~ps). In a large system with $N$ condensates, during one global cycle every node therefore executes
$
N\times\tau/\tau_{\rm rad}\sim 10^5$
local ``spin flips,'' achieving massively parallel calculation within a single nanosecond which is far beyond the reach of CMOS annealers, whose updates must be serialized across time-multiplexed cores (see also Table 1 of \cite{wang2025efficient}). This platform combines ultrafast local physics with extreme parallelism, as all $N$ condensates evolve simultaneously during each feedback interval, and supports non-planar, high-connectivity graphs via free-space optics without the $N^2$ inflation of spins  and accompanying $N^{2-5}$ slowdown in the optimal annealing time that plagues planar embeddings of dense graphs \cite{konz2021embedding}. By introducing mirrors or on-chip delay lines at different distances one can further realize a spectrum of interaction latencies from nearest-neighbor interactions at $\sim10$ps to off-chip long-range links at $\sim1$~ns, providing hierarchical delays that benefit oscillator-based neural networks by combining rapid local consensus with slower global coordination. In combination, these attributes position mirror-mediated polariton networks as an attractive hardware substrate for large-scale analog optimizers, reservoir computers, and physically trained neuromorphic processors.

\end{document}